\documentclass[aps,floatfix,superscriptaddress,reprint,10pt,prl]{revtex4-1}
\usepackage{bbm}
\usepackage{bm}
\usepackage{amsmath}
\usepackage{amssymb}
\usepackage{empheq}
\usepackage{graphicx}
\usepackage{mathrsfs}
\usepackage{amsfonts}
\usepackage{amsthm}
\usepackage{color}
\usepackage{bigints}
\usepackage{txfonts}
\usepackage{hyperref}
\usepackage{color}

\hypersetup{
     unicode=false,               pdftoolbar=true,             pdfmenubar=true,             pdffitwindow=false,          pdfstartview={FitH},         pdftitle={My title},         pdfauthor={Author},          pdfsubject={Subject},        pdfcreator={Creator},        pdfproducer={Producer},      pdfkeywords={keyword1} {key2} {key3},      pdfnewwindow=true,           colorlinks=false,            linkcolor=red,               citecolor=green,             filecolor=magenta,           urlcolor=cyan           }
\makeatletter \tolerance = 10000 \tolerance = 10000
\makeatother

\begin{document}

\title{Spin-orbit related power-law dependence of the diffusive conductivity on
the carrier density in disordered Rashba two-dimensional electron systems}
\author{Weiwei Chen}
\affiliation{Hefei National Laboratory for Physical
Sciences at the Microscale $\&$ Synergetic Innovation
Center of Quantum Information and Quantum Physics,
University of Science and Technology of China, Hefei, Anhui 230026, China}
\author{Cong Xiao}
\thanks{Corresponding author. E-mail: congxiao@utexas.edu}
\affiliation{Department of Physics, The University of Texas at Austin, Austin, Texas 78712, USA}
\author{Qinwei Shi}
\affiliation{Hefei National Laboratory for Physical
Sciences at the Microscale $\&$ Synergetic Innovation
Center of Quantum Information and Quantum Physics,
University of Science and Technology of China, Hefei, Anhui 230026, China}
\author{Qunxiang Li}
\thanks{Corresponding author. E-mail: liqun@ustc.edu.cn}
\affiliation{Hefei National Laboratory for Physical
Sciences at the Microscale $\&$ Synergetic Innovation
Center of Quantum Information and Quantum Physics,
University of Science and Technology of China, Hefei, Anhui 230026, China}

\date{\today}

\begin{abstract}
By using the momentum-space Lanczos recursive method which considers rigorously all
multiple-scattering events, we unveil that the non-perturbative disorder
effect has dramatic impact on the charge transport
of a two-dimensional electron system with Rashba spin-orbit coupling in
the low-density region. Our simulations find a power-law
dependence of the dc longitudinal conductivity on the carrier density, with
the exponent linearly dependent on the Rashba spin-orbit strength but
independent of the disorder strength. Therefore, the classical charge transport
influenced by complicated multiple-scattering processes also shows the
characteristic feature of the spin-orbit coupling. This highly unconventional
behavior is argued to be observable in systems with tunable carrier density
and Rashba splitting, such as the LaAlO$_{3}$/SrTiO$_{3}$ interface,
the heterostructure of Rashba semiconductors bismuth tellurohalides and the
surface alloy Bi$_x$Pb$_y$Sb$_{1-x-y}$/Ag(111).
\end{abstract}

\maketitle

\emph{{\color{blue} Introduction.}}---Spin-orbit coupling underlies numerous fascinating phenomena
in the field of spintronics \cite{Zutic}, such as the spin and anomalous Hall effects \cite{Sinova,Nagaosa}, current-induced spin polarization \cite{Edelstein,Inoue}, and spin-orbit torque \cite{Manchon2019}. Recent studies concerning the interplay between spin-orbit coupling and disorder scattering have successfully described the spin and anomalous Hall effect \cite{Sinova,Nagaosa} in the high carrier density regime. In contrast, how the spin-orbit coupling affects the classical charge transport properties of materials especially in the low charge density regime, such as longitudinal conductivity and Lorentz-force induced Hall effect, is still fuzzy.

Recently, unconventional behaviors of classical charge transport in the two dimensional electronic systems (2DES) with linear Rashba spin-orbit coupling \cite{Nitta,Koo,Manchon} have begun to be uncovered \cite{Brosco,Cong,Ando,Cong-JPCM,Hutchinson,Culcer,Xiao2017,Rashba2007}.
For instance, the Hall coefficient deviates considerably from $1/ne$
in the low-density region ($n<n_{0}$) \cite{Ando,Cong-JPCM}. Here $n$ is the
electron density, {and $n_{0}=m^{2}\alpha _{R}^{2}/(\pi \hbar ^{4})$ is} the
electron density when the Fermi level locates at the Dirac point of the
Rashba system, with $\alpha _{R}$ the Rashba spin-orbit coefficient and $m$
the effective mass. Besides, the longitudinal diffusive conductivities as a
function of $n$ differ significantly between the high-density ($n\geq n_{0}$)
and low-density regions, as shown in the Boltzmann transport theory \cite{Brosco,Cong}:
\begin{equation}
\frac{\sigma}{\sigma_0} =\left\{
\begin{array}{lr}
1, & n\geq n_{0}; \\
\frac{1}{2}(\frac{n^2}{n_{0}^2}+\frac{n^{4}}{n_{0}^{4}}), & n<n_{0}.%
\end{array}%
\right.   \label{Eq:Boltzmann}
\end{equation}%
Here $\sigma _{0}=n_0e^2\tau_0/m$ denotes the conductivity at the Dirac point and {$\tau _{0}=\hbar ^{3}/(mn_{i}\tilde{V}_{0}^{2})$} is the elastic
scattering time, where $\tilde{V}_{0}$ and $n_{i}$ denote, respectively, the
scattering strength and the impurity concentration of Gaussian white-noise
disorder. This formula shows that the diffusive conductivity of classical charge transport is highly
sensitive to the spin-orbit coupling strength in the low-density region.

When the Fermi energy is close to the band edge, however, due to long-wavelength potential fluctuation, previous intensive studies in the absence of spin-orbit coupling confirmed that multiple scatterings off many impurity centers play a dominated role to determine the localized density of states
and invalidate the coherent-potential approximation \cite{Halperin,Galstyan,Wall,Economou,Economou2,Ping}.
As is well known, the presence of spin-orbit coupling which breaks the spin rotational invariance, can transform the orthogonal universality classes into symplectic universality classes and makes the two-dimensional electronic states resilient to the localization \cite{Evangelou,Asada,gorso1,wei}. The mobility edge even locates below the unperturbed band edge in the weak disorder regime.
Therefore, how is the diffusive conductivity in spin-orbit coupled systems influenced by the multiple-scattering is still an open question. In particular, it is of much interest whether the conductivity in this case still shows unconventional characteristic features of the spin-orbit coupling.

A recent work by using the T-matrix approximation predicted plateaus of the
conductivity in the ultra-low-density case of the Rashba system \cite%
{Hutchinson}. The T-matrix approximation only takes into account
multiple scatterings off every single impurity center, but neglects those
off a set of impurities. As a result, it cannot reproduce \cite%
{Hutchinson,Onoda2008} the disorder-induced smooth tail of the density of
states near the band edges, which is however a basic experimental fact \cite{Kato,Kareh}.
Therefore, a more reasonable non-perturbative method is necessary to
inspect the novel transport behavior resulting from multiple-scattering events.

In this work, we simulate the diffusive conductivity of a Rashba 2DES based on the
Kubo formula combined with the Green's function obtained from the Lanczos
recursive method in momentum space. For this purpose, our study focuses on the strong spin-orbit coupling system in the presence of weak-potential disorder, so that even the states a little below the band edge are guaranteed to be extended \cite{wei}.
Our numerical method takes into account rigorously
all multiple-scattering events \cite{Zhu1,Zhu2,Fu}.
We find that in the low-density region the multiple-scattering events lead to an unconventional
power-law dependence of the conductivity $\sigma $ on the electron density:
\begin{equation}
\frac{\sigma}{\sigma_0} =A(\frac{n}{n_{0}})^{\nu },
\label{Eq:Exact}
\end{equation}%
with $A$ a coefficient independent of the electron density.
Our simulation displays that the exponent can be fitted as
\begin{equation}
\nu =-1.56\alpha/t+1.66,  \label{exponent}
\end{equation}
which does not depend on the electron density or disorder strength, but is
linearly related to the spin-orbit strength $\alpha$.

\emph{{\color{blue} Preliminaries.}}---
In the calculation, we simulate the real material by a nearest-neighbor
tight-binding (TB) Hamiltonian on a square lattice,
\begin{equation}
\begin{aligned}
H=2t\sum_{i}c_{i\sigma'}^{\dagger}c_{i\sigma'}-\sum_{\langle
i,j\rangle \sigma'
\sigma''}V_{i\sigma',j\sigma''}c_{i\sigma'}^{\dagger}c_{j\sigma''}+h.c..
\end{aligned}
\end{equation}%
Here
\begin{equation}
V_{i,i+\hat{x}}=\frac{1}{2}\left(
\begin{array}{cc}
t & \alpha \\
-\alpha & t%
\end{array}%
\right) \ ,\ V_{i,i+\hat{y}}=\frac{1}{2}\left(
\begin{array}{cc}
t & -i\alpha \\
-i\alpha & t%
\end{array}%
\right),
\label{Eq:2}
\end{equation}
$c_{i\sigma'}^{\dagger }(c_{i\sigma'})$ denotes the creation (annihilation) operator of an electron on site $i$ with spin $\sigma'$, $t$ stands for the nearest-neighbor hopping energy, and $\alpha$ is the spin-orbit strength.
As we have noted, the existence of metallic phase in the low-density 2DES demands a strong spin-orbit coupling \cite{Evangelou,Asada,gorso1,wei}. Therefore, this study focuses on the regime $0.1\leq\alpha/t\leq0.4$. The upper boundary $\alpha/t=0.4$ in fact represents a very strong spin-orbit coupling in real materials \cite{STO2014,Eremeev,Crepaldi} (see below).

Since in a number of systems, such as semiconductor heterostructures \cite{Miller,Culcer,Wu2014}, LaAlO$_{3}$/SrTiO$_{3}$ interface \cite{STO2010,STO2014}, surface of Rashba semiconductors bismuth tellurohalides \cite{Eremeev,Crepaldi,Nagaosa2012,Ye} and surface alloys \cite{Mirhosseini,Gierz,Ast}, both the carrier density (Fermi energy) and the Rashba spin-orbit coupling can be tuned, experimental verification of Eqs. (\ref{Eq:Exact}) and (\ref{exponent}) is feasible.

In order to simplify the calculation of the conductivity in the Kubo formula, in particular the vertex correction to the conductivity bubble diagram, and to compare with the previous results Eq.~(\ref{Eq:Boltzmann}), we map the TB model into the effective continuum Hamiltonian in the low-energy regime as
\begin{equation}
H(\bm{k})=\frac{\hbar ^{2}\bm{k}^{2}}{2m}+\alpha _{R}(\bm{\sigma}\times %
\bm{k})\cdot \hat{\bm{z}},  \label{Hamilton}
\end{equation}%
with $t=\hbar^2/ma^2$ and $\alpha=\alpha_R/a$. Here $a$ denotes the lattice constant, $\bm{k}=(k_{x},k_{y})$
is the 2D wave-vector, $\hat{\bm{z}}$ is a unit vector perpendicular to the 2D plane,
and $\bm{\sigma}$ is the vector of Pauli matrices. The details of the transformation between the TB model and
continuum model are presented in the Supplemental Materials \cite{supp}
(see, also, references \cite{Shen,Grosso,Halperin1965} therein).
Such a mapping also results in the equality $\alpha/t=k_Ra$, where $k_R=m\alpha_R/\hbar^2$
corresponds to the Rashba wave-vector which measures the momentum splitting of the two Rashba sub-bands.
To give a specific example, we consider the surfaces of Bismuth Tellurohalides \cite{Shevelkov,Eremeev},
where $k_{R}\approx 0.05\mathring{\mathrm{A}}$$^{-1}$, $a\approx 4.3\mathring{\mathrm{A}}$ and
hence $\alpha\approx 0.22t$. The two models match well in the low-density regime
when the spin-orbit coupling $\alpha\leq0.4t$. Beyond this value, the mapping from the TB model
to the continuum one gradually fails to work because one can no longer obtain the same dispersions
even at very low energies \cite{supp}.

The eigenfunctions and eigenvalues of $H(\bm{k})$ (Eq.~\ref{Hamilton})
are given respectively by{\ $|\bm{k}s\rangle =\frac{1}{\sqrt{2}}(i,se^{i\theta _{%
\bm{k}}})^{\text{T}}$ and $E_{\bm{k}s}=\frac{\hbar ^{2}\bm{k}^{2}}{2m}%
+s\alpha _{R}|\bm{k}|$}, where $s=\pm 1$ denotes the helicity and $\theta _{%
\bm{k}}$ is defined by $\theta _{\bm{k}}=\arctan
(k_{y}/k_{x})$. The two Rashba bands $E_{\bm{k}s}$ are
approximately linear in the vicinity of the Dirac point $\bm{k}=0$, where
they touch each other. The matrix
\begin{equation}
\begin{aligned} U_{\bm{k}}=\frac{1}{\sqrt{2}}\left(\begin{array}{ccc} i &
i\\ e^{i\theta_{\bm{k}}} & -e^{i\theta_{\bm{k}}}\\ \end{array}\right)
\end{aligned}  \label{Eq:U}
\end{equation}%
\noindent
implements the rotation from the spin to the eigenstate basis.
Besides, the disorder is modeled by the
Gaussian white noise, $\overline{V(\bm{r})V(\bm{r'})}=n_{imp}\tilde{V}%
_{0}^{2}\delta (\bm{r}-\bm{r'}${$)$, where }$\overline{\cdots }$ stands for
averaging over disorder realizations.

Within the linear response the longitudinal diffusive conductivity at
zero-temperature is given by the Kubo formula \cite{Mahan,Bruus}
\begin{equation}
\sigma (E)=\sigma ^{RA}(E)-\sigma ^{RR}(E),  \label{conductivity}
\end{equation}%
where
\begin{equation}
\begin{aligned} \sigma^{RA}(E)=\frac{e^2\hbar}{2\pi}\int\frac{d^2\bm{k}}{(2\pi)^2}{\rm
Tr}[G^{R}(\bm{k},E)v_xG^{A}(\bm{k},E)\tilde{v}_{x}] \end{aligned},
\end{equation}%
\begin{equation}
\begin{aligned} \sigma^{RR}(E)=\frac{e^2\hbar}{2\pi}\int\frac{d^2\bm{k}}{(2\pi)^2}{\rm Re}{\rm
Tr}[G^{R}(\bm{k},E)v_xG^{R}(\bm{k},E)\tilde{v}_{x}] \end{aligned}.
\label{RR}
\end{equation}%
Here Tr represents the trace over helicity $s$, and
\begin{equation}
\begin{aligned} & G(\bm{k},E)=\left(\begin{array}{ccc} g(\bm{k}+,E) & 0\\ 0 &
g(\bm{k}-,E)\\ \end{array}\right) \end{aligned}  \label{Eq:Green}
\end{equation}%
denotes the Green's function of the disordered system in the band-eigenstate
basis with $g(\bm{k}s,E)=(E-E_{\mathbf{k}s}-\Sigma (\bm{k}s,E))^{-1}$ and $%
\Sigma (\bm{k}s,E)$ the self-energy. $A,R$ indicate advanced or retarded
Green's functions. The $x$ component of the velocity operator in the
band-eigenstate basis is given by $v_{x}=\frac{1}{\hbar }(\frac{\hbar
^{2}k_{x}}{m}+\alpha _{R}\cos \theta \sigma _{z}+\alpha _{R}\sin \theta
\sigma _{y}),$ and the vertex function $\tilde{v}_{x}$ can be obtained from
the Bethe-Salpeter equation $\tilde{v}_{x}(\mathbf{k})=v_{x}(\mathbf{k}%
)+n_{imp}\tilde{V}_{0}^{2}\int \frac{d^{2}\mathbf{p}}{4\pi ^{2}}U_{\mathbf{k}%
}^{\dagger }U_{\mathbf{p}}G({\mathbf{p}},E)\tilde{v}_{x}(\mathbf{p})G({%
\mathbf{p}},E)U_{\mathbf{p}}^{\dagger }U_{\mathbf{k}}.\label{vertex}$ Based
on symmetry arguments, it is verified that $\tilde{v}_{x}$ has the same
matrix structure as $v_{x}$, so that the vertex function can be solved as
\begin{equation}
\tilde{v}_{x}=\frac{1}{\hbar }(\frac{\hbar ^{2}k_{x}}{m}+\tilde{\alpha}%
_{R}\cos \theta \sigma _{z}+\tilde{\alpha}_{R}\sin \theta \sigma _{y}),
\end{equation}%
where%
\begin{equation}
\begin{aligned} & \tilde{\alpha}_R=\frac{\alpha_R
+n_{imp}\tilde{V}_0^{2}I_{1}}{1-n_{imp}\tilde{V}_0^{2}I_{2}},\\ & I_1=\int
\frac{d^2\textbf{k}}{4\pi^2} \frac{\hbar^2k}{4m}(g_{+}g_{+}-g_{-}g_{-}), \\
& I_2=\int \frac{d^2\textbf{k}}{4\pi^2}
\frac{1}{4}(g_{+}g_{+}+g_{-}g_{-}+g_{+}g_{-}+g_{-}g_{-}). \end{aligned}
\label{I1I2}
\end{equation}%
with $g_{\pm }=g(\bm{k}\pm ,E)$. Thus the conductivity can be calculated by
Eqs.~(\ref{conductivity}-\ref{I1I2}) with exactly calculated Green's functions.

\emph{{\color{blue} Numerical methods.}}---In our numerical simulation, the
Green's functions $g(\bm{k}s,E)$ of the disordered systems are calculated
using the well-developed Lanczos recursive method \cite{Zhu1,Zhu2,Fu} with the TB model.
We generate the disorder by random on-site
energies with zero mean and $V_{0}^{2}$ variance, where $V_{0}=\tilde{V}%
_{0}/a^{2}$, without loss of generality. The impurity concentration is $%
n_{i}=1/a^{2}$ in following calculation.

The numerical evaluation requires a nonzero broadening (resolution) parameter $\eta\gtrsim\delta E$, where $\delta E$ is the mean level spacing \cite{Ferreira}.
In order to obtain a high energy resolution and also be free from the finite-size errors, we consider a large enough square lattice of
size $L_{x}\times L_{y}=8000\times 8000$ with periodic boundary conditions
in both the $x$ and $y$ directions. Thus, a small artificial parameter $\eta =0.001t$
is used to simulate the infinitesimal imaginary energy in our simulations.
Remarkably, based on the standard Dyson equation $\Sigma (\bm{k}%
s,E)=g_{0}^{-1}(\bm{k}s,E)-g^{-1}(\bm{k}s,E)$, we find the self-energy
function is independent of both $\bm{k}$ and $s$.
\begin{figure}[t]
\begin{center}
\includegraphics[width=0.45\textwidth]{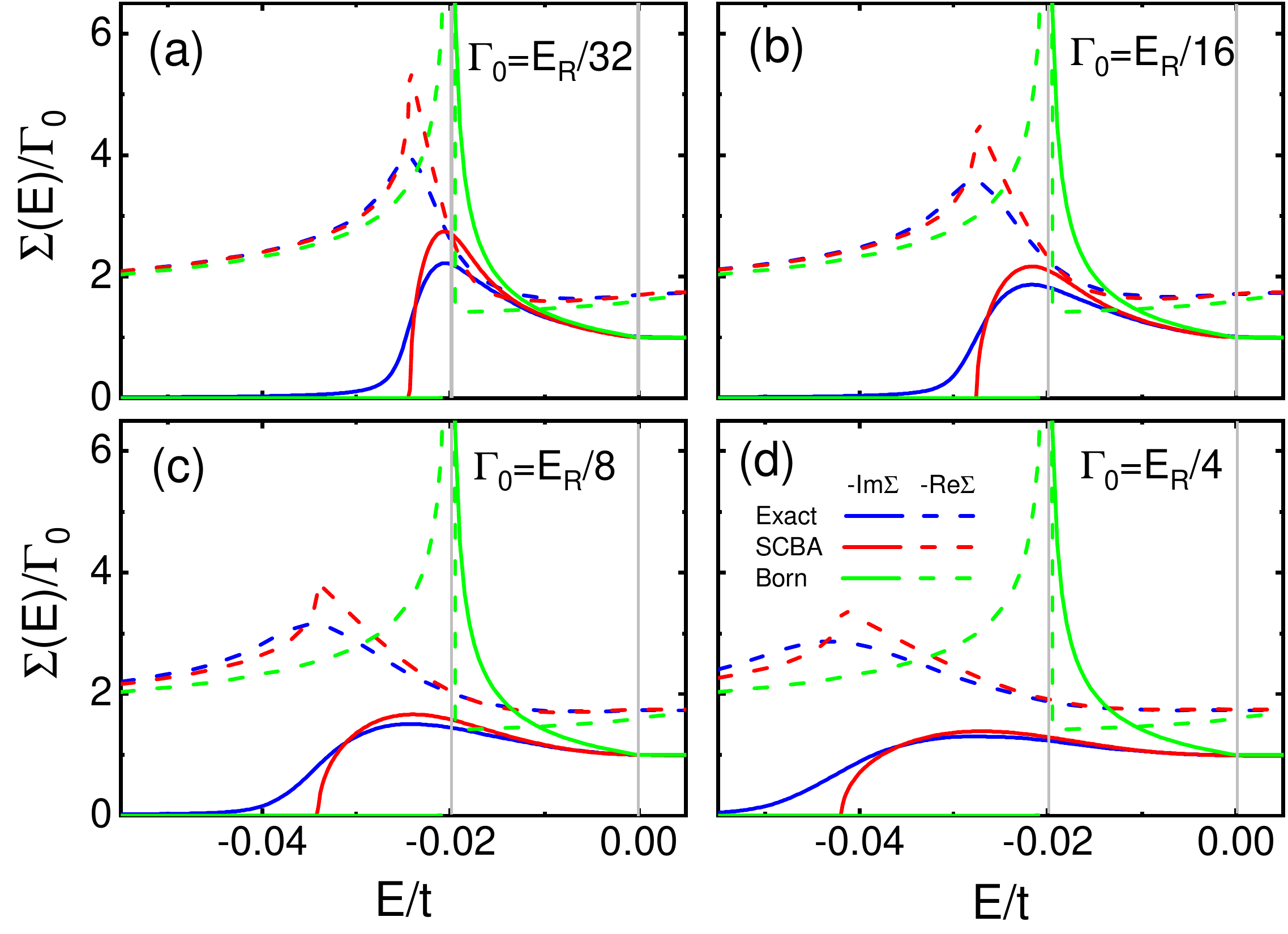}
\end{center}
\par
\vspace{-1em}
\caption{ (Color online) The self-energy function versus energy of the
system with the spin-orbit strength $\protect\alpha /t=0.2$ and the disorder
strength: (a) $\Gamma _{0}=E_{R}/32$, (b) $\Gamma _{0}=E_{R}/16$, (c) $%
\Gamma _{0}=E_{R}/8$, and (d) $\Gamma _{0}=E_{R}/4$. The results calculated
from the exact numerical simulation (blue), the SCBA (red) and the Born
approximation (green) are displayed for comparison. Gray lines locate at the
Dirac point $E=0t$ and band edge of the pure system $E=-E_{R}=-0.02t$. Here $%
\Gamma _{0}=\hbar /2\protect\tau _{0}=V_{0}^{2}/{2t}$ denotes the
disorder-induced band broadening.}
\label{Fig:self-energy}
\end{figure}

Before addressing the transport behaviors, here we show the advantage of our
exact simulation to the self-energy over other methods employed in previous
studies on the Rashba system, including the Born approximation \cite%
{Brosco,Cong}, self-consistent Born approximation (SCBA) \cite{Brosco} and
the T-matrix approximation \cite{Hutchinson,Onoda2008}. The self-energy
produced by the latter two methods are qualitatively similar \cite%
{Hutchinson,Onoda2008}, so we do not show the result of the T-matrix
approximation. In Fig.~\ref{Fig:self-energy} we plot the numerical real and imaginary parts of self-energy as functions of the Fermi energy for different disorder strengthes, and compare them with the results of Born approximation and SCBA.
As expected, the Born's and SCBA's results both work well in the high-density regime, where the perturbation approaches are successful due to the presence of a small parameter expansion in terms of $1/k_Fl$. Here $k_F$ and $l$ denote the Fermi momentum and mean free path, respectively. On the contrary, as $E$ approaches the band edge, $k_Fl\lesssim 1$ brings the system into a totally different regime where the contribution from multiple scattering events plays in important role and the conventional perturbative methods are invalid \cite{Bruus,Imry}. The effects of multiple scattering involving many impurity centers on the self-energy, for instance the the crossing wigwam self-energy diagrams sketched in the Supplemental Materials \cite{supp}, are out of the regime of previous perturbation theories. However, they become important in the strong-scattering case ($k_Fl\lesssim 1$). Thus, the results of the Born approximation and SCBA gradually deviate from our non-perturbative results including all the multiple scattering contributions. Especially, the tail of the imaginary part of the SCBA self-energy vanishes sharply, contrary to the smooth tail in our numerical simulation. Such a sharp reduction behavior may lead to some unphysical behaviors, for example, the upturn of mobility near the band edge in the previous SCBA calculation \cite{note}.

It is worthwhile to note that the character of the imaginary part of the self-energy
obtained by our simulation is consistent with the smooth tail of the
experimental density of states of the Rashba-type spin-split states near the
conduction band bottom, such as the surface state of Bi/Ag(111) \cite%
{Kato,Kareh}. This agreement indicates that our simulation indeed gives a
reasonable account for the multiple-scattering effects in the low-density
region of Rashba systems.
\begin{figure}[t]
\begin{center}
\includegraphics[width=0.45\textwidth]{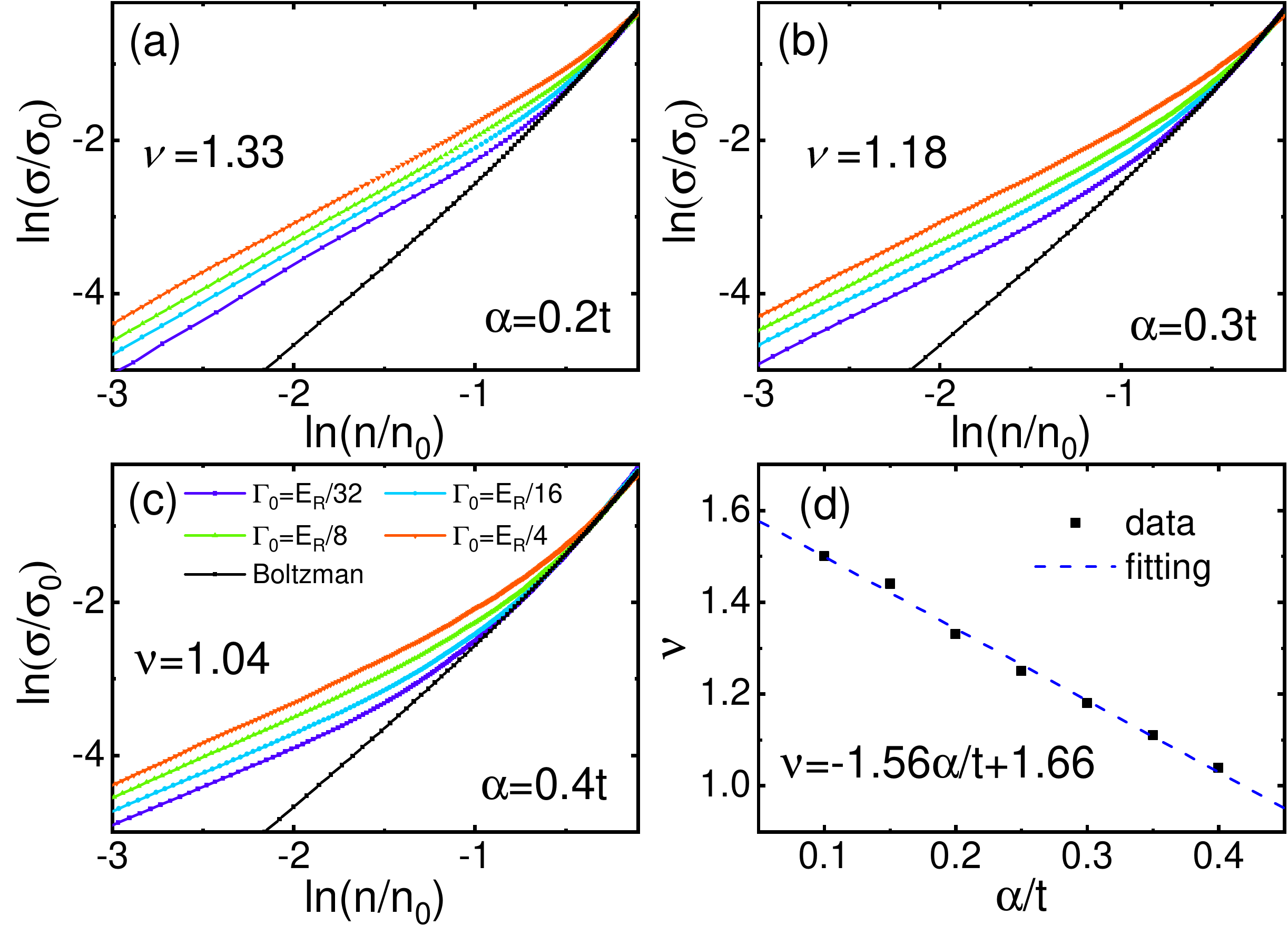}
\end{center}
\par
\vspace{-1em}
\caption{ln$(\protect\sigma /\protect\sigma _{0})$ vs ln$(n/n_{0})$ for
spin-orbit strength (a) $\protect\alpha =0.2t$, (b) $\protect\alpha =0.3t$
and (c) $\protect\alpha =0.4t$. In the low-density regime, the numerical results are described
by Eq. (\protect\ref{Eq:Exact}) with $\protect\nu $ independent of both the
carrier density and random disorder strength. The Boltzmann analytical
result is plotted for comparison. (d) The slope $\protect\nu $ of ln$(%
\protect\sigma /\protect\sigma _{0})$ vs ln$(n/n_{0})$ in the low-density
region as a function of spin-orbit strength $\protect\alpha /t$.}
\label{fig:conductivity}
\end{figure}

\emph{{\color{blue} Spin-orbit related power-law conductivity.}}---The qualitative difference between the self-energies produced by our simulation
and by the SCBA or the T-matrix approximation suggests that our method may
demonstrate some transport behaviors unprecedented in previous theoretical
researches of 2D Rashba systems \cite{Brosco,Cong,Hutchinson}. Our
simulation supports this speculation by finding an emergent power-law
dependence of the diffusive conductivity on the carrier density [Eq.~(\ref%
{Eq:Exact})] in the low-density region.
The curves of the conductivity versus the carrier density $n$ for different
spin-orbit strengths ($\alpha /t=0.2,0.3$ and $0.4$) are displayed in the
log-log plots Fig.~\ref{fig:conductivity}(a), (b) and (c), compared with the
Boltzmann analytical formula [Eq.~(\ref{Eq:Boltzmann})]. In the low-density
regime our results deviate significantly from the analytical solution \cite{note}.

\begin{figure}[t]
\begin{center}
\includegraphics[width=0.45\textwidth]{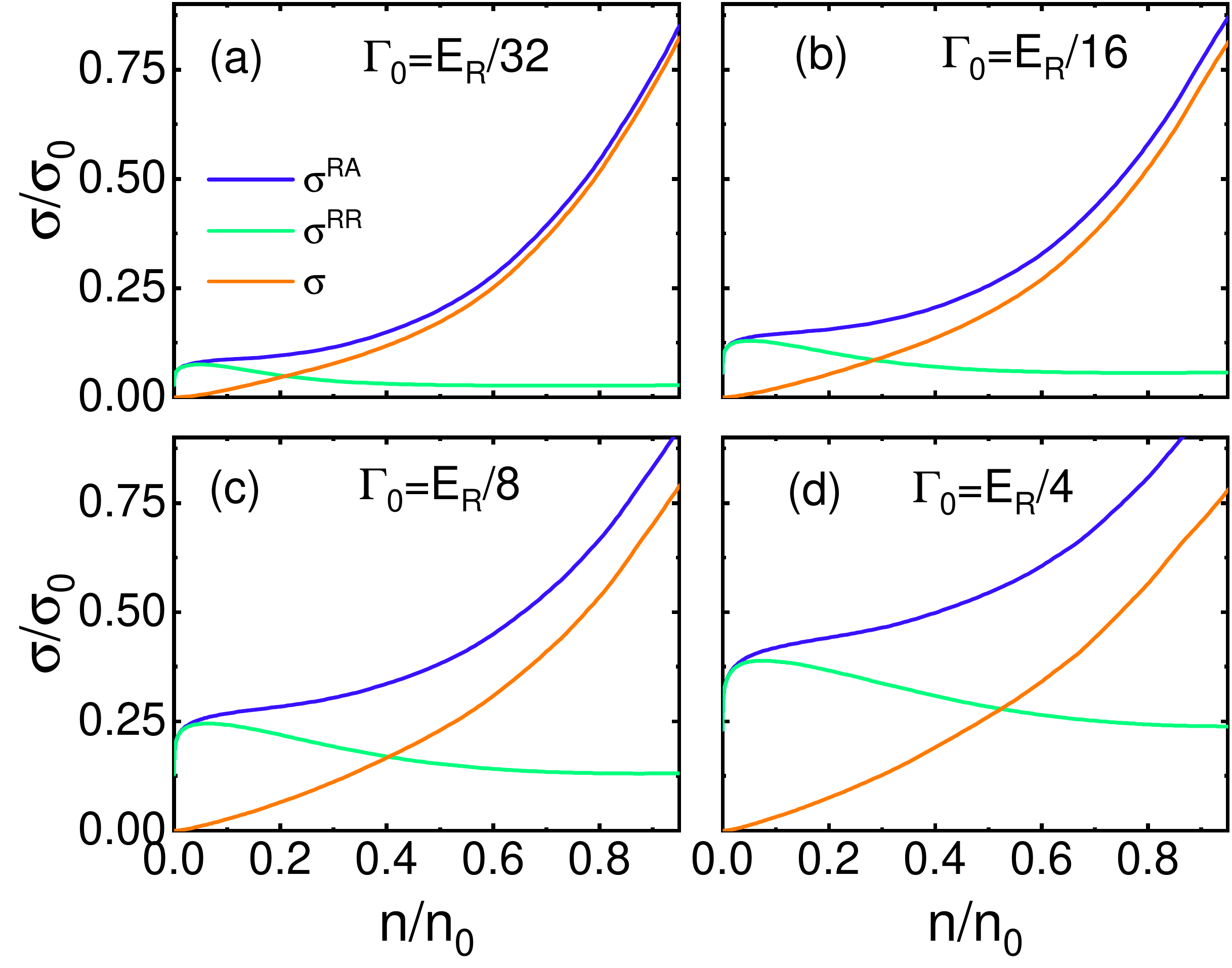}
\end{center}
\par
\vspace{-1em}
\caption{Different contributions to the conductivity as a function of the
charge density for systems with $\protect\alpha /t=0.2$ and disorder
strengthes (a) $\Gamma _{0}=E_{R}/32$, (b) $\Gamma _{0}=E_{R}/16$, (c) $%
\Gamma _{0}=E_{R}/8$ and (d) $\Gamma _{0}=E_{R}/4$. }
\label{fig:RR}
\end{figure}

In the multiple-scattering dominated regime the
curves of $\ln (\sigma /\sigma _{0})$ vs $\ln (n/n_{0})$ in Fig.~\ref%
{fig:conductivity}(a), (b) and (c)\ are mostly linear. This observation
inspires us to use the power-law formula [Eq.~(\ref{Eq:Exact})] to fit the
results, where the exponent $\nu $ is independent of the carrier density. As
shown in Fig.~\ref{fig:conductivity}(a), (b) and (c), the curves
corresponding to different disorder strengths $\Gamma _{0}/E_{R}=1/4$, 1/$8$%
, 1/$16$ and 1/$32$ (defined in the caption of Fig.~\ref{Fig:self-energy})
for a fixed spin-orbit strength are parallel to each other in the linear
regime. This means that the exponent $\nu $ in Eq.~(\ref{Eq:Exact}) is also
independent of the random disorder strength.

When presenting the values of $\nu $ for different
spin-orbit strengths in the same plot, Fig.~\ref{fig:conductivity}(d), we
find that the exponent $\nu $ is linearly dependent on the spin-orbit
strength $\alpha $. Fitting the data, we obtain the linear scaling Eq.~(\ref%
{exponent}). This equation indicates that, the charge transport
influenced by complicated multiple-scattering processes also shows the
characteristic feature of the spin-orbit coupling.
The deep understanding for the underlying physical mechanism
leading to this unconventional relation is not clear at the present stage and is
beyond the scope of our numerical study. More theoretical efforts are called
for in the future. Here we just numerically find this relation, which can be
experimentally tested as a transport indicator of multiple-scattering.

Another remark here is that, the factor $A$ in Eq.~(\ref{Eq:Exact}) is
dependent on both the disorder and spin-orbit strengths. The $A$-$V_{0}$
curves for different spin-orbit strengths are shown in the Supplemental
Materials \cite{supp}. In the considered regime $0.1t\leq\alpha\leq0.4t$ we can approximately fit $A$ as
$A(\alpha/t,V_{0}/t)=0.47\left( \alpha /t\right) ^{-1.43}V_{0}/t+0.03\left( \alpha/t\right) ^{-1.1}$.


\emph{{\color{blue} Conclusion and discussion.}}---In conclusion, we showed that the multiple-scattering events play an important role in determining
both the quasi-particle and transport properties of the low-density Rashba 2DES.
Our simulations uncover a power-law dependence of the dc
conductivity on the electron density with the exponent linearly dependent on the
spin-orbit strength but independent of the disorder strength.

To provide some clues in understanding the unconventional transport behavior
described by Eqs.~(\ref{Eq:Exact}) and (\ref{exponent}), we stress here the
relevance of the $\sigma ^{RR}$ term [Eq.~(\ref{RR})]. Theoretically, this
term can be neglected in the Boltzmann regime where the $\sigma ^{RA}$ term
yields the quantitatively similar result to Eq.~(\ref{Eq:Boltzmann}).
Hence, when the non-Boltzmann power-law
conductivity emerges instead of the Boltzmann formula, the $\sigma ^{RR}$
term is anticipated to be important. In Fig.~\ref{fig:RR}, the contributions
from the $\sigma ^{RR}$ and $\sigma ^{RA}$ terms are shown separately for
the case of $\alpha /t=0.2$. In combination with Fig.~\ref{fig:conductivity}%
(a), we find that the power-law [Eq.~(\ref{Eq:Exact})] holds perfectly when $%
\sigma ^{RR}\geq \sigma ^{RA}/3$.

So far we have assumed scalar (spin independent) short-range scatterers.
Here we note that the short-range disorder can be classified into three types according to the spin dependence: spin independent, spin conserved and spin-flipped. In the Supplemental Materials \cite{supp} we display $\ln(\sigma/\sigma_0)$ versus $\ln(n/n_0)$ in the cases of the other two types of disorder: the spin conserved disorder $V_1=V_1(\bm{r})\sigma_z$ and the spin-flipped disorder $V_2=\bm{V}_2(\bf{r})\cdot\bm{\sigma}$. Here $\bm{V}_2(\bm{r})$ is a in-plane vector, and both $V_1(\bm{r})$ and $\bm{V}_2(\bm{r})$ are random with zero mean and $V^2_0$ variance. In these two cases the exponent of the conductivity power-law can be fitted respectively by the linear relations $\nu=-1.36\alpha/t+1.41$ and $\nu=-2.42\alpha/t+1.83$. Therefore, we find that the linear relation in Eq.~(\ref{exponent}) holds for each type of short-range disorder, with the slope and intercept constants depending on the type of disorder. We also find that the Eq.~(\ref{exponent}) is independent of the impurity concentration \cite{supp}.

Lastly, we suggest some experimental systems where our simulation results
can be potentially observed.
First, attention can be paid to the Rashba 2DESs in heterostructures,
due to the tunability of the Rashba effect by an external electric field,
such as the one formed at the LaAlO$_{3}$/SrTiO$_{3}$ interface \cite{STO2010,STO2014}
where $k_{R}\approx 0.08\mathring{\mathrm{A}}$$^{-1}$ ($a\approx 2.5\mathring{\mathrm{A}}$, $\alpha /t\approx 0.2$) in the absence of the external electric field.
Besides, the Rashba 2DESs in the heterostructures formed by n-type
polar semiconductors bismuth tellurohalides are also compelling candidates \cite{Wu2014}.
Meanwhile, the Rashba 2DESs appearing near the surface of bismuth tellurohalides \cite{Eremeev,Crepaldi,Nagaosa2012,Ye} can be considered for experiment as well,
such as that in the surface states of Bismuth Tellurohalides \cite{Shevelkov,Eremeev}
which arises in the bulk-gap region with $k_{R}\approx 0.05\mathring{\mathrm{A}}$$^{-1}$ ($a\approx 4.3\mathring{\mathrm{A}}$, $\alpha /t\approx 0.22$).
Furthermore, it has been reported that in the surface alloy Bi$_x$Pb$_y$Sb$_{1-x-y}$/Ag(111) \cite{Mirhosseini,Gierz,Ast} the Fermi energy and Rashba splitting
can be independently tuned through the concentrations $x$ and $y$.
This system may be another good platform to verify our finding.


\begin{acknowledgments}
We thank Yunshan Cao, Peng Yan and Chen Wang for making the cooperation possible.
C.X. is indebted to Yunke Yu for her mental support in the beginning days of the cooperation.
This work is supported by the National Key Research \& Development Program of China (Grants No. 2016YFA0200604)
and the National Natural Science Foundation of China (Grants No. 21873088, and 11874337).
C.X. is supported by NSF (EFMA-1641101) and Welch Foundation (F-1255).
\end{acknowledgments}

\end{document}